\documentclass[11pt]{article}
\usepackage[]{amscd,amsfonts,amsthm}
\newtheorem{basicnotion}{Basic Notion}
\newtheorem{theorem}{Theorem}
\newtheorem{definition}{Definition}

\newtheorem{proposition}{Proposition}

\usepackage[psamsfonts]{amssymb}
\newcommand{\real}{{\mathbb R}}
\newcommand{\bd}{\begin{definition}}
\newcommand{\ed}{\end{definition}}
\newcommand{\bp}{\begin{proposition}}
\newcommand{\ep}{\end{proposition}}
\newcommand{\be}{\begin{equation}}
\newcommand{\ee}{\end{equation}}
\newcommand{\bea}{\begin{eqnarray}}
\newcommand{\eea}{\end{eqnarray}}
\newcommand{\ba}{\begin{array}}
\newcommand{\ea}{\end{array}}
\begin{document}

\title{Being and Change: Foundations of a Realistic Operational Formalism\footnote{Published as: D. Aerts, ``Being and change:
foundations of a realistic operational formalism", in {\it Probing the Structure of Quantum Mechanics: Nonlinearity, Nonlocality,
Computation and Axiomatics}, eds. D. Aerts, M. Czachor and T. Durt, World Scientific, Singapore (2002).}}

\author{Diederik Aerts}

\date{}
\maketitle

\centerline{Center Leo Apostel (CLEA) and}
\centerline{Foundations of the Exact Sciences
(FUND),}
\centerline{Brussels Free
University, Krijgskundestraat 33,}
\centerline{1160 Brussels,
Belgium.}
\centerline{diraerts@vub.ac.be}

\begin{abstract}
\noindent
The aim of this article is to represent the general description of an entity by means of its states,
contexts and properties. The entity that we want to describe does not necessarily have to be a physical entity, but
can also be an entity of a more abstract nature, for example a concept, or a cultural artifact, or the mind of a
person, etc..., which means that we aim at very general description. The effect that a context has on the state of
the entity plays a fundamental role, which means that our approach is intrinsically contextual. The approach is
inspired by the mathematical formalisms that have been developed in axiomatic quantum mechanics, where a specific
type of quantum contextuality is modelled. However, because in general states also influence context -- which is not
the case in quantum mechanics -- we need a more general setting than the one used there. Our focus on context as a
fundamental concept makes it possible to unify `dynamical change' and `change under influence of measurement', which
makes our approach also more general and more powerful than the traditional quantum axiomatic approaches. For this
reason an experiment (or measurement) is introduced as a specific kind of context. Mathematically we introduce  a
state context property system as the structure to describe an entity by means of its states, contexts and properties.
We also strive from the start to a categorical setting and derive the morphisms between state context property
systems from a merological covariance principle. We introduce the category {\bf SCOP} with as elements the state
context property systems and as morphisms the ones that we derived from this merological covariance principle. We
introduce property completeness and state completeness and study the operational foundation of the formalism.
\end{abstract}

\section{Introduction}
We put forward a formalism that aims at a general description of an entity under influence of a context. The approach
followed in this article is a continuation of what has been started in \cite{Aerts1999a,Aerts1999b}. Meanwhile some new elements
have come up, and as a consequence what we present here deviates in some ways from what was elaborated in
\cite{Aerts1999a,Aerts1999b}. In \cite{Aerts1999a} the formalism is founded on
the basic concepts of states, experiments and outcomes of experiments. One of
the aims in \cite{Aerts1999a} is to generalize older approaches
\cite{Piron1976,Aerts1981,Aerts1982,Aerts1983a,Aerts1983b,Piron1989,Piron1990} by incorporating experiments with more than two outcomes
from the start. The older approaches are indeed founded on the basic concept of yes/no-experiment, also called test,
question or experimental project. An experiment with more than two outcomes is described by the set of yes/no-experiments that can be
formed from this experiment by identifying outcomes in such a way that, after identification, only two outcomes remain. The shift
away from the old approach with yes/no-experiments was inspired by the extra power, at first sight at least, that formalisms that
start from experiments with more than two outcomes as basis posses \cite{RF1975,RF1978,FR1981,RF1981,FPR1983,RF1983}. Meanwhile,
after \cite{Aerts1999a,Aerts1999b} had been published, we have identified a fundamental problem that exists in approaches
that start with experiments with more than two outcomes. 

\subsection{The Subexperiment Problem in Quantum Mechanics} \label{subsec:subexprob}

The problem that we have identified is the following. Suppose that in quantum mechanics one considers a subexperiment of an
experiment. Then it is always the case that the subexperiment changes the state of the entity under study in a different way than
the experiment does. If however a subexperiment is defined as the experiment where some of the outcomes are
identified, the subexperiment must change the state in the same way as the original experiment. This means that in quantum mechanics
subexperiments do not arise through an identification process on the outcomes of an experiment. This also means that the general
scheme inspired from probability theory, as for example in
\cite{RF1975,RF1978,FR1981,RF1981,FPR1983,RF1983}, where outcomes are taken as basic concepts and events and experiments are defined by
means of their identifying sets of outcomes, does not work for quantum mechanics. The subexperiments that can be fabricated in these
type of approaches are not the subexperiments that one encounters in quantum mechanics.  

When we were writing \cite{Aerts1999a}, we
started to get aware of this fundamental problem. We were very amazed that to the best of our knowledge nobody else working
in quantum axiomatics, ourselves included, seemed to have noticed this before. It
puts a fundamental limitation on the approaches where one starts with experiments with more outcomes and deduces the subexperiments, for
example the yes/no-experiments, by a procedure of identification of outcomes, as in the approach presented in
\cite{Aerts1999a}. In \cite{Aerts1999b} we have avoided the problem by again working with yes/no-experiments as basic concepts. In this
article we will introduce experiments, and hence also yes/no-experiments, in another way, not avoiding the problem, but tackling it
head on. In \cite{ADH2002} we analyze in detail the subexperiment problem in quantum mechanics.

\subsection{Applying the Formalism to Situations Outside the Microworld} \label{subsec:applicationotherfields}
The development of the formalism has always been grounded by applying it to specific examples of physical entities in specific
situations. Already in the early years it became clear that the formalism, that originally was developed with the aim of providing
a realistic operational axiomatic foundation for quantum mechanics, can be applied to describe entities that are not part of the
micro-physical region of reality. Since the formalism delivers a general description of an entity, it is a priori not mysterious
that it can be applied to entities that are part of other regions of reality than the microworld. The study of such entities is
interesting on its own and also can shed light on the nature of quantum entities. In this sense we presented a first macroscopic
mechanical entity entailing a quantum mechanical structure in \cite{Aerts1985,Aerts1986,Aerts1987}. More specifically this entity
represents a model for the spin of a spin${1 \over 2}$ quantum entity.

The model puts forward an explanation of the quantum probabilities as
due to the presence of fluctuations on the interaction between the measuring apparatus and the physical entity under consideration.
We have called this aspect of the formalism the `hidden measurement
approach' \cite{AA1997b,AACDDV1997,Aerts1998a,AS1998,AADL1999,ACS1999,AS2002,DDH2002,DBMRL2002}. The name refers to the fact that
the presence of these fluctuations can be interpreted as the presence of hidden variables in the measuring apparatus (hidden
measurements) instead of the presence of hidden variables in the state of the physical entity, which is what traditional hidden
variable theories suppose. Concretely it means that, when a measurement is performed, the actual measuring process that makes occur
one of the outcomes is deterministic once it starts. But each time a measurement is repeated, which is necessary to obtain the
probability as limit of the relative frequency, a new deterministic process starts. And the presence of fluctuations on the
interaction between measurement apparatus and physical entity are such that this new repeated measurement can give rise
deterministically to another outcome than the first one. The quantum probability arises from the statistics of how each time again
a new repeated measurement can give rise to another outcome, although once started it evolves deterministically. 

As mentioned already, concrete realizable mechanical models that expose this situation were built, where it is possible to `see'
how the quantum structure arises as a consequence of the presence of fluctuations on the interaction between measurement and
entity \cite{Aerts1985,Aerts1986,Aerts1987,Aerts1988,AS1998,AADL1999,AS2002,Aerts1991,Aerts1993,Aerts1994}. Specifically these
macroscopic mechanical models that give rise to quantum structure inspired us to try out whether the formalism could be applied to
entities in other fields of reality than the micro-physical world.

A first application was elaborated with the aim of describing the situation of decision making. More concretely we made a model 
\cite{AA1994,AA1997} for an opinion pole situation, where also persons without opinion can be described, and the effect of the questioning
during the opinion pole itself can be taken into account.

We also constructed a cognitive situation where Bell inequalities are violated because of the presence of a nonKolmogorovian
probability structure in the cognitive situation \cite{AABG2000a}. We worked out in detail a description of the liar paradox by means of the
Hilbert space structure of standard quantum mechanics \cite{ABS1999a,ABS1999b}.

More recently our attention has gone to studies of cultural evolution \cite{GA2000,Gabora2001}, cognitive science, more
specifically the problem  of the representation of concepts \cite{GA2000,Gabora2001}, and biology with the aim of
elaborating a global evolution theory \cite{Gabora2001,AG2002}. One of the important insights that has grown
during the work on \cite{GA2000}, and the work found in Liane Gabora's doctoral thesis \cite{Gabora2001}, is that experiments (or
measurements) can be considered to be contexts that influence the state of the entity under consideration in an indeterministic
way, due to fluctuations that are present on the interaction between the context and the entity. As we knew already from our work
on the hidden measurement approach, such indeterministic effects give rise to a generalized quantum structure for the description
of the states and properties of the entity under study. This makes it possible to classify dynamical evolution for classical and
quantum entities and change due to measurements on the same level: dynamical change being deterministic change due to
dynamical context and measurement change being indeterministic change due to measurement context, the indeterminism finding its
origin on the presence of fluctuations on the interaction between context and entity.

In the present article it is the first time that we present the formalism taking into account this new insight that unifies dynamical
change with measurement change. As will be seen in the next sections, this makes it necessary to introduce some fundamental changes
in the approach.

\subsection{States Can Change Context} \label{subsec:stateschangecontexts}

Our investigations in biological and cultural evolution \cite{GA2000,Gabora2001,AG2002} have made it clear that in general change
will happen not only by means of an influence of the context on the state of the entity under study, but also by means of the
influence of the state of the entity on the context itself. Interaction between the entity and the context gives rise to a change
of the state of the entity, but also to a change of the context itself. This possibility was not incorporated in earlier versions
of the formalism. The version that we propose in this paper takes it into account from the start, and introduces an
essential reformulation of the basic setting. In this paper, although we introduce the possibility of change of the context under
influence of the state from the start, we will not focus on this aspect. We refer to \cite{GA2000,Gabora2001} for a more detailed
investigation of this effect, and more specifically to \cite{AG2002} for a full elaboration of it.

\subsection{Filtering Out the Mathematical Structure} \label{subsec:filteringmath}
Already in the last versions of the formalism \cite{Aerts1999a,Aerts1999b} the power of making a good distinction between the
mathematical aspects of the formalism and its physical foundations had been identified. Let us explain more concretely what we
mean. In the older founding papers
\cite{Piron1976,Aerts1981,Aerts1982,Aerts1983a,Aerts1983b,Piron1989,Piron1990}, although the physical foundation of the
formalism is defined in a clear way, and the resulting mathematical structures are treated rigorously, it is not always
clear what are the `purely mathematical' properties of the structures that are at the origin of the results. That is the reason that
in more recent work on the formalism we have made an attempt to divide up the physical foundation and the resulting mathematical
structure as much as possible. We first explain in which way certain aspects of the mathematical structure
arise from the physical foundation, but then, in a second step, define these aspects in a strictly mathematical way, such that
propositions and theorems can be proven, `only' using the mathematical structure without physical interpretation. Afterwards, the results of
these propositions and theorems can then be interpreted in a physical way again. This not only opens the way for mathematicians to start
working on the structures, but also lends a greater axiomatic strength to the whole approach on the fundamental level. More concretely, the
mathematical structure of a state property system is the the structure to be used to describe a physical entity by means of its states and
properties\cite{Aerts1999a,Aerts1999b,ACVVVS1999}. This step turned out to be fruitful from the start, since we could prove that
a state property system as a mathematical structure is isomorphic to a closure space \cite{Aerts1999a,Aerts1999b,ACVVVS1999}. This
means that the mathematics of closure spaces can be translated to the mathematics of state property systems, and in this sense
becomes relevant for the foundations of quantum mechanics. The step of dividing up the mathematics from the physics in a
systematic way also led to a scheme to derive the morphisms for the structures that we consider from a covariance principle
rooted in the relation of a subentity to the entity of which it is a subentity\cite{Aerts1999a,ACVVVS1999}. This paved the
way to a categorical study of the mathematical structures involved, which is the next new element of the
recent advances that we want to mention.

\subsection{Identifying the Categorical Structure} \label{subsec:categoricalaspects}
Not only was it possible to connect with a state property system a closure space in an isomorphic way, but, after we had
introduced the morphisms starting from a merological covariance principle, it was possible to prove that the category of state
property systems and their morphisms, that we have named {\bf SP}, is equivalent with the category of closure
spaces and continuous functions, denoted by {\bf Cls} \cite{Aerts1999a,ACVVVS1999}. More specifically we could prove that {\bf
SP} is the amnestic modification of {\bf Cls} \cite{ACVVVS2001}. Meanwhile these new element in the approach have lead to
strong results. 

It could be proven that some of the axioms of axiomatic quantum
mechanics \cite{Piron1964,Aerts1981,Aerts1982} correspond to separation properties of the
corresponding closure spaces \cite{VanSteirteghem1998}. More concretely, the axiom of state
determination in a state property system \cite{Aerts1999a} is equivalent with the
$T_0$ separation axiom of the corresponding closure space \cite{VanSteirteghem1998,VanSteirteghem2000}, and the axiom of
atomicity in a state property system \cite{Aerts1999a} is equivalent with the $T_1$ separation axiom of the corresponding
closure space
\cite{VanderVoorde2000,VanderVoorde2001}. More recently it has been shown
that `classical properties' \cite{Aerts1981,Aerts1983a,Piron1989,Piron1990} of the state property system correspond to
clopen (open and closed) sets of the closure space \cite{ADVV2001,ADVV2002,AD2002}, and, explicitly making use of the categorical
equivalence, a decomposition theorem for a state property system into its nonclassical components can be proved that
corresponds to the decomposition of the corresponding closure space into its connected components
\cite{ADVV2001,ADVV2002,AD2002}.

\subsection{The Introduction of Probability} \label{subsec:probability}

In \cite{Aerts1999a,Aerts1999b} we introduce for the first time probability on the same level as the other concepts, such
as states, properties and experiments. Indeed, in the older approaches, probability was not introduced on an equally
fundamental level as it is the case for the concepts of state, property and experiment. 

What probability fundamentally tries to do is to provide a `measure' for the uncertainty that is present in the situation of entity
and context. We had introduced probability in a standard way in \cite{Aerts1999a,Aerts1999b} by means of a measure on the interval
$[0, 1]$ of the set $\real$ of real numbers. Meanwhile however it has become clear that probability should better be introduced in a
way that is quite different from the standard way. We have analyzed this problem in detail in \cite{Aerts2002}. We come to the
conclusion that it is necessary to define a probability not as an object that is evaluated by a number in the interval $[0, 1]$, as
it is the case in standard probability theory, but as an object that is evaluated by a subset of the interval $[0, 1]$. We have called this
type of generalized probability -- standard probability theory
is retrieved when the considered subsets of the interval $[0, 1]$ are the singletons -- a `subset probability'. Although a lot of work
is still needed to make the subset probability into a full grown probability theory, we will take it in account for the elaboration of the
formalism that we propose in this paper.

\subsection{How We Will Proceed}
We will take into account all the aspects mentioned in sections \ref{subsec:subexprob}, \ref{subsec:applicationotherfields},
\ref{subsec:stateschangecontexts}, \ref{subsec:filteringmath},
\ref{subsec:categoricalaspects} and \ref{subsec:probability} for the foundations of the formalism that we introduce in this paper.
We also try to make the paper as self contained as possible, such that it is not necessary for the reader to go through all the
preceding material to be able to understand it.

For sake of completeness we mention other work that has contributed to advances in the
approach that is less directly of importance for what we do in this paper, but shows how the approach is developing into other
directions as well
\cite{dEmma1980,Daniel1982,CDPGN1988,CN1991,CN1992,CN1993,Moore1995,Aerts1996a,Aerts1996b,ADH1996,ACDH1997,ACDV1997a,ACDV1997b,Aerts1998b,ACS1998,Stubbe1998,Aerts1999c,ABG1999,AC1999,Piron1999,CS1999a,CS1999b,Aerts2000,ABG2000b,AVS2000,CM2000,CS2000,Aerts2001,Valckenborgh2001,Smets2001}.

\section{Foundations of the Formalism} 

In this section we introduce the basic ingredients of the formalism. Since we want to be able to apply the formalism to many
different types of situations the basic ingredients must be sufficiently general. The strategy that we follow consists of
describing more specific situations as special cases of the general situation. The primary concept that we
consider is that of an {\bf entity}, that we will denote by $S$. Such an entity
$S$ can for example be a cat, or a
genome, or cultural artifact, such as a building, or an abstract idea, or a
mind of a person, or a stone, or a quantum particle, or a 
fluid etc \ldots

\subsection{States, Contexts and Change}

At a specific moment, an entity, {\it is} in a specific {\bf
state}. The state represents
{\it what the entity is and how it reacts to different contexts} at that
moment. For example, a cat can be awake
or asleep, this are two possible states of the cat. The second basic concept that we consider is that of a {\bf context}. A context
is a part of the outside reality of the entity that influences the entity in such a way that its state is changed. If the cat
is asleep, and confronted with a
context of heavy noise, it is probable that it will wake up.
The context `heavy noise' changes then the state `cat is asleep' into the state
`cat is awake'.

For a specific entity $S$ we denote the states of this entity
by $p, q, r, s, \dots$ and the contexts
that the entity can be confronted with by $e, f, g, h, \ldots$. The set of all
relevant states of an entity $S$ we denote by
$\Sigma$ and the set of all relevant contexts by ${\cal M}$.

Let us express the basic situation that we consider. An entity $S$,
in a state $p$, interacts with a context $e$. In general, the
interaction between the entity and the
context causes a change of the state of the
entity and also a change of the context. Sometimes there will be no change
of state or no change of context: this
situation we consider as a special case of the general situation; we can
for example call it a situation of zero change.
Sometimes the entity will be destroyed by the context with which it
interacts. A context that destroys the entity 
provokes a change of the state that the entity is in before it interacts
with this context, but this change is so strong that
most of the characteristic properties of the entity are destroyed, and
hence the remaining part of reality will not be
identified any longer as the entity. To be able to express the destruction
within our formalism we introduce a special state $0
\in \Sigma$. If a context changes a specific state $p \in
\Sigma$ to the state $0 \in
\Sigma$ it means that the entity in question is destroyed by the
context\footnote{The state $0$ is not really a
state in the proper sense, it is introduced specifically to take into
account the possibility for a context to destroy
the entity that we consider.}. We denote by $\Sigma_0$ the set of all
relevant states of the entity
$S$ without the state of destruction $0$. Let us introduce the basic notions in a formal way.

\begin{basicnotion} [State, Context and Change]
For an entity $S$ we introduce its set of relevant states
$\Sigma$ and its set of relevant contexts
${\cal M}$. States are denoted by $p, q, r, s \in \Sigma$ and contexts are
denoted by $e, f,
g, h \in {\cal M}$. The state $0 \in \Sigma$ is the state that expresses
that the entity is destroyed. By $\Sigma_0$
we denote the set of states without the state $0$ that represent the
destroyed entity. The situation where the entity $S$ is
in a state
$p
\in
\Sigma$ and under influence of a context $e \in {\cal M}$ will in general
lead to a new situation where the entity is in a
state $q \in \Sigma$ and the context is $f \in {\cal M}$. The transition
from the couple $(e, p)$ to the couple $(f, q)$ we
call the change of the entity in state $p$ under influence of the context
$e$.
\end{basicnotion}

\subsection{Probability}
Probability theory has been developed to get a grip on indeterminism. Usually a
probability is defined by means of a measure on the interval $[0, 1]$ of the set of real numbers $\real$. As we mentioned already
in section \ref{subsec:probability}, we are not convinced that the standard way to introduce probability is the good way for our
formalism. We have analyzed this problem in detail in \cite{Aerts2002} and use the results obtained there.

\bd [Probability of Change] \label{def:probability}
Consider an entity $S$ with a set of states $\Sigma$ and a set of
contexts ${\cal M}$. We introduce the function
\bea
\mu: {\cal M} \times \Sigma \times {\cal M} \times \Sigma &\rightarrow& {\cal P}([0, 1]) \\
(f, q, e, p) \mapsto \mu(f, q, e, p)
\eea
where $\mu(f, q, e, p)$ is the probability for the couple $(e, p)$ to change to the couple $(f, q)$. ${\cal P}([0, 1])$ is the set of
all subsets of the interval $[0, 1]$. This means that we evaluate the probability by a subset of $[0, 1]$ and not a number of $[0, 1]$
as in standard probability theory.
\ed
\noindent
This is a generalization of standard probability theory that we retrieve when all the considered
subsets are singletons. It will become clear in the following what are the advantages of this generalization as is also explained in
detail in \cite{Aerts2002}. We remark that if the probability that we introduce would be a traditional probability, where all the $\mu(f,
q, e, p)$ are singletons, we would demand the following type of property to be satisfied:
\be
\sum_{f \in {\cal M}, q \in \Sigma}\mu(f, q, e, p) = 1
\ee
expressing the fact that the couple $(e, p)$ is changed always to one of the couples $(f, q)$. This rule is usually referred to as the
sum rule for probability. For a subset probability the sum rule is much more complicated. We refer to
\cite{Aerts2002} for a more complete reflection on this matter, but admit that the matter has not been solved yet. In the
course of this article we will make use of the sum rule for subset probabilities only to express that indeed a couple
$(e, p)$ is always changed to a couples $(f, q)$.

\subsection{States and Properties}
A property is something that the
entity `has' independent of the type of context that the entity is confronted with. That is the reason why we consider properties to be
independent basic notions of the formalism. We denote properties by $a, b, c,
\ldots$ and the set of all relevant properties of an entity by ${\cal L}$. 

For a state we require that an entity is in a
state at each moment . The state represents the reality of the entity at that moment. Properties are elements of this reality. This means
that an entity
$S$ in a specific state $p
\in
\Sigma$ has different properties that are actual. Properties that are actual for the entity being in a specific
state can be potential for this entity in another state.
\begin{basicnotion} [Property]
For an entity $S$, with set of states $\Sigma$ and set of contexts ${\cal M}$, we introduce its set of properties. A
property can be actual for an entity in a specific state and potential for this entity in another state. We denote properties by $a,
b, c, \ldots$, and the set of properties of $S$ by ${\cal L}$.
\end{basicnotion}
\noindent
We introduce some additional concepts to be able to express the basic situation that we want to consider in relation with the state
and the properties of an entity. Suppose that the entity
$S$ is in a specific state
$p
\in
\Sigma$. Then some of the properties of $S$ are actual and some are
not and hence are potential. This means that with each state $p \in
\Sigma$ corresponds a set of actual properties, subset of ${\cal L}$.
This defines a function $\xi : \Sigma \rightarrow {\cal
P}({\cal L})$, which maps each state $p \in \Sigma$ to the set $\xi(p)$ of properties that are actual in this
state. Introducing this function makes it possible to
replace the statement `property $a \in {\cal L}$ is actual for the
entity $S$ in state $p \in \Sigma$' by `$a \in \xi(p)$'.
\par
Suppose now that for the entity $S$ a specific
property $a \in {\cal L}$ is actual. Then this entity is in a certain
state $p \in \Sigma$ that makes $a$ actual. With each
property
$a
\in {\cal L}$ we can associate the set of states that make this
property actual, i.e. a subset of $\Sigma$. This
defines a function $\kappa : {\cal L} \rightarrow {\cal P}(\Sigma)$,
which makes  each property $a \in {\cal L}$ correspond to the set of
states $\kappa(a)$ that make this property actual. Again we can replace the statement
`property $a \in {\cal L}$ is actual if the entity $S$ is in state $p
\in \Sigma$' by the set theoretical expression `$p \in \kappa(a)$'. Let us introduce these notions in a formal way.
\bd [Aristotle Map]
Consider an entity $S$, with set of states $\Sigma$, set of contexts ${\cal M}$ and set of properties ${\cal L}$. We define a
function $\xi: \Sigma \rightarrow {\cal P}({\cal L})$ such that $\xi(p)$ is the set of all properties that are actual if the entity
is in state $p$. We call $\xi$ the Aristotle map \footnote{We introduce the name Aristotle map for this function as a homage to Aristotle,
because he was the first to consider the set of actual properties as characteristic for the state of the considered entity.} of the entity
$S$.
\ed
\bd [Cartan Map]
Consider an entity $S$, with set of states $\Sigma$, set of contexts ${\cal M}$ and set of properties ${\cal L}$. We define a
function $\kappa: {\cal L} \rightarrow {\cal P}(\Sigma)$ such that $\kappa(a)$ is the set of all states that make the property $a$
actual. We call $\kappa$ the Cartan map\footnote{The name Cartan map for this function was introduced in earlier formulations of
the formalism \cite{Aerts1981,Aerts1982} as an homage to Eli Cartan, who for the first time considered the state
space as the fundamental structure in the case of classical mechanics}.
\ed
\noindent
By introducing the Aristotle map and the Cartan map we can express the basic situation we want to consider for states and
properties as follows:
\begin{equation} \label{statprop}
 a \in \xi(p) \Leftrightarrow p \in \kappa(a) \Leftrightarrow  a \ {\rm is\ actual\ for\ } S\ {\rm in\ state\ } p
\end{equation}
If the state $p \in \Sigma$ of an entity is changed under influence of a context $e \in {\cal M}$ into a state $q \in \Sigma$, then
the set $\xi(p)$ of actual properties in state $p$ is changed into the set $\xi(q)$ of actual properties in state $q$. Contrary to
the state of an entity, a property is not changed under influence of the context. What can be changed is its status of actual or
potential. This change of status under influence of the context is monitored completely by the change of
state by this context.

\subsection{Covariance and Morphisms}
We derive the morphism of our structure by making use of a merological covariance principle. What we mean
is that we will express for the situation of an entity and subentity the covariance of
the descriptions, and in this way derive the morphisms of our structure.

Consider two entities $S$ and $S'$ such that $S$ is a subentity of $S'$. In that case, the following three natural
requirements should be satisfied:

\smallskip
\noindent i) If the entity $S'$ is in a state $p'$ then the state $m(p')$ of
$S$ is determined. This defines a function $m$ from the set of states of $S'$ to the set of states of $S$. 

\smallskip
\noindent ii) If we consider a context $e$ that influences the entity $S$, then this context also influences the entity $S'$. This
defines a function $l$ from the set of contexts of $S$ to the set of contexts of $S'$. 

\smallskip
\noindent
iii) When we consider a property $a$ of $S$, then this corresponds to a property
$n(a)$ of $S'$, which  is the same property, but now considered as a property of the big entity $S'$. This defines a function $n$ of
${\cal L}$ to ${\cal L}'$.

\smallskip
\noindent
iv) We want $e$ and $l(e)$ to be two descriptions of the `same' context, once
considered as a context that influences $S$ and once considered as a context that influences $S'$. This means
that when we consider how $e$ influences $m(p')$, this is the same physical process as when we consider how $l(e)$ influences $p'$,
with the only difference that the first time it is considered within the description of the subentity $S$ and the second time
within the description of the big entity $S'$. In a similar way we want property $a$ to behave in the same way towards the
states of $S$ as property $n(a)$ behaves towards the states of $S'$. As a consequence the following covariance principles hold:

\bea
\mu(f, m(q'), e, m(p')) &=& \mu'(l(f), q', l(e), p') \\
a \in \xi(m(p')) &\Leftrightarrow& n(a) \in \xi'(p')
\eea
We have now everything at hand to define a morphism of our structure.

\bd [Morphisms]\label{def:morphism}
Consider two entities $S$ and $S'$, with sets of states $\Sigma$, $\Sigma'$, sets of contexts ${\cal M}$, ${\cal M}'$ and sets of
properties ${\cal L}$ and ${\cal L}'$, such that probability is defined as in definition 
\ref{def:probability}. We say that the triple (m, l, n) is a morphism if $m$ is a function:
\begin{equation}
m: \Sigma' \rightarrow \Sigma
\end{equation}
$l$ is a function:
\begin{equation}
l: {\cal M} \rightarrow {\cal M}'
\end{equation}
and $n$ is a function:
\be
n: {\cal L} \rightarrow {\cal L}'
\ee
such that for $p' \in \Sigma'$, $e \in {\cal M}$ and $a \in {\cal L}$ the following holds:
\bea
\mu(f, m(q'), e, m(p')) &=& \mu'(l(f), q', l(e), p') \label{cov:01}\\
a \in \xi(m(p')) &\Leftrightarrow& n(a) \in \xi'(p') \label{cov:02}
\eea
\ed
\noindent
Requirements (\ref{cov:01}) and (\ref{cov:02}) are the covariance formula's that characterize the morphisms
of our structure. 

\subsection{The Category of State Context Property Systems and Their Morphisms}

We define now the mathematical structure that we need to describe an entity by means of its states, context and properties, in a
purely mathematical way, so that the structure can be studied mathematically. 

\bd [The Category SCOP]
A state context property system \\ $(\Sigma, {\cal M}, {\cal L}, \mu, \xi)$, consists of three sets $\Sigma$, ${\cal M}$, ${\cal
L}$ and two functions
$\mu$ and
$\xi$, such that
\bea
\mu: {\cal M} \times \Sigma \times {\cal M} \times \Sigma \rightarrow {\cal P}([0, 1]) \\
\xi: \Sigma \rightarrow {\cal P}({\cal L})
\eea
The sets $\Sigma$, ${\cal M}$ and ${\cal L}$, play the role of the set of states, the set of contexts, and the set of properties of
an entity $S$. The function
$\mu$ describes transition probabilities between couples of contexts and states, while the function $\xi$ describes the
sets of actual properties for the entity $S$ being in different states.

Consider two state context
property systems $(\Sigma, {\cal M}, {\cal L}, \mu, \xi)$ and \\ $(\Sigma', {\cal M}', {\cal L}', \mu', \xi')$. A morphism is a
triple of functions $(m, l, n)$ such that:
\bea
m: \Sigma' &\rightarrow& \Sigma \\
l: {\cal M} &\rightarrow& {\cal M}' \\
n: {\cal L} &\rightarrow& {\cal L}
\eea
and the following formula's are satisfied:
\bea
\mu(f, m(q'), e, m(p')) &=& \mu'(l(f), q', l(e), p') \label{scom:06} \\
a \in \xi(m(p')) &\Leftrightarrow& n(a) \in \xi'(p') \label{scom:07}
\eea
We denote the category consisting of state context property systems and their morphisms by {\bf SCOP}.
\ed

\section{Classical and Quantum Entities}
As we mentioned, our formalism must be capable to describe different
forms of change that are encountered
in nature for different types of entities. Let us consider some examples to show how this happens.

\subsection{Classical and Quantum Dynamical Change} \label{subsec:clasquant}
In this section we outline in which way the
deterministic evolution of classical physical
entities described by the dynamical laws of classical physics and the
deterministic evolution of
quantum entities described by the Schr\"odinger
equation are described in the formalism. 

For a classical mechanical entity in a certain state $p$ under a
context $e$ change is
deterministic. A specific couple
$(e, p)$ changes deterministically into another couple $(f, q)$. Quantum entities undergo two types of
change. One of them,
referred to as the dynamical change and described by the
Schr\"odinger equation, is also deterministic. The other one, referred to as collapse and described by von Neumann's projection
postulate, is indeterministic. We consider in this section only the deterministic Schr\"odinger type of change for a quantum
entity.

Let us consider first a concrete example of the classical mechanical change. Place a charged iron ball in in a magnetic
field. The
ball will be influenced by the context which is the magnetic field and start to move according to the
classical laws of
electromagnetism. Knowing the context at $t_0$ and the state of
the entity at $t_0$ we can predict with certainty
the state of the entity at
$t_1$, $t_2 \dots$. If we consider the influence of the magnetic
field from time $t_0$ to
time $t_1$ as one context $e(t_0 \mapsto t_1)$ and the influence of the
magnetic field from time
$t_1$ till time $t_2$ as another context $e(t_1 \mapsto t_2)$, then the
whole dynamical evolution can
be seen as the change monitored continuously by the set of contexts
$e(t_1 \mapsto t)$. All
these contexts change the state of the entity in a
deterministic way. 

The deterministic evolution of a
quantum entity from time $t_1$
to time $t$, described in the standard quantum formalism by means of the
Schr\"odinger equation, can be represented
in our formalism in the same way, as the change
monitored continuously by the set of contexts $e(t_1 \mapsto
t)$.

Classical and quantum
dynamical evolution are both characterized as a continuous change by
an infinite set of contexts
that all change the state of the entity in a deterministic way. We can specify this
situation in the general formalism. To be able to express in a nice way the specific characteristics of quantum and classical
entities, we introduce the concept of range.

\bd [Range of a Context for a State]
Consider a state context property system $(\Sigma, {\cal M}, {\cal L}, \mu, \xi)$ describing an entity $S$. For $p \in \Sigma$ and
$e \in {\cal M}$ we introduce:
\be 
R(e, p) = \{q\ \vert\ q \in \Sigma,\ \exists\ f \in {\cal M}\ {\rm such\ that}\ \mu(f, q, e, p) \not= \{0\}\}
\ee
and call $R(e, p)$ the range of $e$ for $p$.
\ed
\noindent
The range of a context for a state is the set of states that this state can be changed to under influence of this context. We can now
easily define what is a deterministic context to a state.

\bd [Deterministic Context to a State]
Consider a state context property system $(\Sigma, {\cal M}, {\cal L}, \mu, \xi)$ describing an entity $S$. We call a context $e \in
{\cal M}$ a deterministic context to a state $p \in \Sigma$ if $R(e, p) = \{q\}$. We call
$q$ the image of the state $p$ under the context $e$.
\ed
\noindent
If $e$ is a deterministic context to the state $p$, then the state
$p$ is changed
deterministically to its image state $q$ under influence of the context $e$.

\bd [Range of a Context]
Consider a state context property system $(\Sigma, {\cal M}, {\cal L}, \mu, \xi)$ describing an entity $S$. For $e \in {\cal M}$ we
introduce:
\be
R(e) = \cup_{p \in \Sigma}R(e, p)
\ee
and call $R(e)$ the range of the context $e$.
\ed
\noindent
The range of a context $e$ is the set of states that is reached by changes provoked by $e$ on any state of the entity.  We can of
course also define the range of a state for a context. It is the set of contexts that this context can change to under influence of
this state.
\bd[Range of a State for a Context]
Consider a state context property system $(\Sigma, {\cal M}, {\cal L}, \mu, \xi)$ describing an entity $S$. For
$e \in {\cal M}$ and $p \in \Sigma$ we introduce:
\be 
R(p, e) = \{f\ \vert\ f \in {\cal M},\ \exists\ q \in \Sigma\ {\rm such\ that}\ \mu(f, q, e, p) \not= \{0\}\}
\ee
and call $R(p, e)$ the range of $p$ for $e$.
\ed
\noindent
This
makes it easy to define what is a deterministic state to a context.
\bd [Deterministic State to a Context]
Consider a state context property system $(\Sigma, {\cal M}, {\cal L}, \mu, \xi)$ describing an entity $S$. We call a state $p \in
\Sigma$ a deterministic state to a context $e \in {\cal M}$ if $R(p, e) = \{f\}$. We call
$f$ the image of the context $e$ under the state $p$.
\ed
\noindent
If $p$ is a deterministic state to the context $e$ then the context
$e$ is changed
deterministically to its imagine context $f$ under influence of the state $p$.

\bd [Range of a State]
Consider a state context property system $(\Sigma, {\cal M}, {\cal L}, \mu, \xi)$ describing an entity $S$. For $p \in \Sigma$ we
introduce:
\be
R(p) = \cup_{e \in {\cal M}}R(p, e)
\ee
and call $R(p)$ the range of the state $p$.
\ed
\noindent
The range of a state $p$ is the set of contexts that is reached by changes provoked by $p$ on any context of the entity.

\bd [Deterministic Context State Couple]
Consider a state context property system $(\Sigma, {\cal M}, {\cal L}, \mu, \xi)$ describing an entity $S$. We call the context $e \in
{\cal M}$ and the state $p \in \Sigma$ a deterministic context state couple, if $e$ is a deterministic context to $p$, and $p$ is a
deterministic state to $e$. Then $(e, p)$ changes deterministically to $(f, q)$. We call $(f, q)$ the image of $(e, p)$.
\ed

\bd [Deterministic Context]
Consider a state context property system $(\Sigma, {\cal M}, {\cal L}, \mu, \xi)$ describing an entity $S$. We say that a context $e
\in {\cal M}$ is a
deterministic context if it is a deterministic context to each state $p \in
\Sigma$.
\ed
\noindent
For quantum entities only the contexts that generate the dynamical
change of state described by the
Schr\"odinger equation are deterministic contexts. For classical entities
all contexts are deterministic.

\bd [Deterministic State]
Consider a state context property system $(\Sigma, {\cal M}, {\cal L}, \mu, \xi)$ describing an entity $S$. We say that a state $p
\in \Sigma$ is a deterministic state if it is a deterministic state to any context $e \in
{\cal M}$.
\ed
\noindent
For quantum mechanics as well as for classical physics all states are deterministic states. In physics we indeed do not in general
consider an influence of the state on the context. 

We have
now all the material to
define what is in our general formalism a
d-classical entity.

\bd [D-Classical Entity]
Consider a state context property system $(\Sigma, {\cal M}, {\cal L}, \mu, \xi)$ describing an entity $S$. We say that
the entity $S$ is a d-classical entity if all its states and contexts are deterministic.
\ed
\noindent
The reason why we call such an entity a d-classical entity and not just a classical entity is that our definition only demands the
classicality of the entity towards the change of state that can be provoked by a context. There exist other
possible forms of classicality, for example towards the type of properties that an entity can have. This is
the reason the we prefer to call the type of classicality that we introduce here d-classicality. In \cite{AD2002} the structure
related to d-classical entities is analyzed in detail. 

We note that in our formalism it are the
deterministic contexts that produce an entity to
behave like the physical entities behave under dynamical evolution, whether
this is the evolution of classical
physical entities under any kind of context, or the evolution of quantum
physical entities described by the
Schr\"odinger equation.

\subsection{Quantum Measurement Contexts}

In the foregoing section we have seen that the context that gives rise to a quantum evolution is a deterministic context. For
quantum entities there are also contexts that originate in the measurement. These contexts are not deterministic. Let us see how
their actions fit into the formalism. Consider a quantum entity in a state
$p$ represented by a unit vector $u(p)$ of a complex Hilbert space ${\cal H}$. A measurement context $e$ in quantum mechanics is
described by a self-adjoint operator $A(e)$ on this Hilbert space. In general, a self-adjoint operator has a spectrum that consists
of a point-like part and a continuous part. In the point-like part of the spectrum, the state $p$ is transformed into one of the
eigenstates represented by the eigenvectors corresponding to the points of the point-like part of the spectrum of $A(e)$. For the continuous
part of the spectrum the situation is somewhat more complicated. There are no points as outcomes here, but only intervals. To such an
interval corresponds a unique projection operator of the spectral resolution of $A(e)$, and the state $p$ is then projected by this
projection onto a vector of the Hilbert space, which represents that state after the measurement.

\subsection{Cultural Change and the Human Mind}

Cultural change with the human mind as generating entity is very complex. Here states as well as context will in general not be
deterministic. How the formalism can be applied there has been studied in \cite{GA2000,Gabora2001}. More specifically a situation
describing `the invention of the torch' has been modelled in
\cite{Gabora2001}. We will not consider this type of change in more detail in the present article and refer to \cite{AG2002} for a
formal approach concentrated on this case.

\subsection{Eigen States and Eigen Contexts}
There is a type of determinism of a context towards a state and of a state towards a context that we want to consider specifically, namely
when there is no change at all.

\bd [Eigenstate of a Context]
Consider a state context property system $(\Sigma, {\cal M}, {\cal L}, \mu, \xi)$ describing an entity $S$. A state $p \in \Sigma$ is
called an eigenstate of the context $e \in {\cal M}$ if the context $e$ is deterministic to the state $p$ and the image of $p$ under
$e$ is $p$ itself.
\ed

\bp
Consider a state context property system $(\Sigma, {\cal M}, {\cal L}, \mu, \xi)$ describing an entity $S$. A state $p \in \Sigma$
is an eigenstate of the context $e \in {\cal M}$ iff
\be
R(e, p) = \{p\}
\ee
\ep

\bd [Eigencontext of a State]
Consider a state context property system $(\Sigma, {\cal M}, {\cal L}, \mu, \xi)$ describing an entity $S$. A context $e \in {\cal M}$ is
called an eigencontext of the state $p \in \Sigma$ if the state $p$ is deterministic to the context $e$ and the image of $e$ under $p$ is $e$
itself.
\ed

\bp
Consider a state context property system $(\Sigma, {\cal M}, {\cal L}, \mu, \xi)$ describing an entity $S$. A context $e \in {\cal
M}$ is an eigencontext of the state $p \in \Sigma$ iff
\be
R(p, e) = \{e\}
\ee
\ep
\noindent
The name eigenstate has been taken from quantum mechanics, because indeed if a quantum entity is in an eigenstate of
the operator that represents the considered measurement, then this state is not changed by the context of this measurement. Experiments in
classical physics are observations and hence do not change the state of the physical entity.

\section{Pre-Order Structures}

Step by step we introduce additional structure in our formalism. As much as possible we introduce this structure in an
operational way, meaning that we analyze carefully what is the meaning of the structure that we introduce and how it is connected
with reality. In this section we limit ourselves to the identification of a natural pre-order structure on the set of states and on
the set of properties. 

\subsection{State and Property Implication and Equivalence}
Before introducing the state and property implications that will form pre-order relations on $\Sigma$ and ${\cal L}$, let us
first define what is a pre-order relation on a set.

\bd [Pre-order Relation and Equivalence]
Suppose that we have a set $Z$. We say that $<$ is a pre-order
relation on $Z$ iff for $x, y, z \in Z$ we have:
\be
\begin{array}{l}
x < x \\
x < y \ {\rm and}\ y < z \Rightarrow x < z
\end{array}
\ee
For two elements $x , y \in Z$ such that $x < y$ and $y < x$
we denote $x \approx y$ and we say that $x$ is equivalent to $y$.
\ed
\noindent
There exist natural `implication relations' on $\Sigma$ and on ${\cal L}$. If the situation is such that if `$a \in
{\cal L}$ is actual for
$S$ in state
$p \in \Sigma$' implies that `$b \in {\cal L}$ is actual for $S$ in
state $p \in \Sigma$' we say that property $a$ `implies' property $b$. If the situation is such that `$a \in
{\cal L}$ is actual for
$S$ in state
$q \in \Sigma$' implies that `$a \in {\cal L}$ is actual for $S$ in
state $p \in \Sigma$' we say that the state $p$ implies the
state $q$. Let us introduce these two implications in a formal way.
\bd [State Implication and Property Implication]
Consi\-der a state context property system $(\Sigma, {\cal M}, {\cal L}, \mu, \xi)$ describing an entity $S$. For $a, b \in
{\cal L}$ we introduce:
\be \label{ordprop}
a < b \Leftrightarrow \kappa(a) \subset \kappa(b)
\ee
and we say that $a$ `implies' $b$. For $p, q \in \Sigma$ we introduce:
\be \label{ordstat}
p < q \Leftrightarrow \xi(q) \subset \xi(p)
\ee
and we say that $p$ `implies' $q$ \footnote{Remark that the state
implication and property implication are not defined in a
completely analogous way. Indeed, then we should have written $p
< q \Leftrightarrow \xi(p) \subset \xi(q)$. That we have chosen
to define the state implication the other way around is because
historically this is how intuitively is thought about
states implying one another.}.
\ed
\noindent
It is easy to verify that the implication relations that we
have introduced are pre-order relations.
\bp
Consider a state context property system $(\Sigma, {\cal M}, {\cal L}, \mu, \xi)$ describing an entity $S$.
Then $\Sigma, <$ and ${\cal L}, <$ are pre-ordered sets.
\ep
\noindent
We can prove the following:
\bp
Consider a state context property system $(\Sigma, {\cal M}, {\cal L}, \mu, \xi)$ describing an entity $S$.
 (1) Suppose that $a, b \in {\cal L}$ and $p \in \Sigma$. If $a \in
\xi(p)$ and $a < b$, then $b \in \xi(p)$. (2) Suppose that $p, q
\in
\Sigma$ and $a \in {\cal L}$. If $q \in \kappa(a)$ and $p < q$
then $p \in \kappa(a)$.
\ep \label{statprop04}
\noindent
Proof: (1) We have $p \in \kappa(a)$ and $\kappa(a) \subset \kappa(b)$.
This proves that $p \in \kappa(b)$ and hence $b \in \xi(p)$. (2) We
have
 $a \in \xi(q)$ and $\xi(q) \subset \xi(p)$ and hence $a \in \xi(p)$.
This shows that
$p \in \kappa(a)$. \qed

\medskip
\noindent
It is possible to prove that the morphisms of the category {\bf SCOP} conserves the two implications.
\bp
Consider two state context property systems $(\Sigma, {\cal M}, {\cal L}, \mu, \xi)$ and $(\Sigma', {\cal M}', {\cal L}', \mu',
\xi')$ describing entities $S$ and $S'$, and a morphism $(m, l, n)$ between these two state context. For $p', q' \in \Sigma'$ and
$a, b \in {\cal L}$ we have:
\bea
p' < q ' &\Leftrightarrow& m(p') < m(q') \label{form:impl01} \\
a < b &\Leftrightarrow& n(a) < n(b)  \label{form:impl02}
\eea 
\ep
\noindent
Proof: Suppose that $p' < q'$. This means that $\xi'(q') \subset \xi'(p')$. Consider $a \in \xi(m(q'))$. Using \ref{scom:07}
implies that $n(a) \in \xi(q')$. But then $n(a) \in \xi(p')$ and again using \ref{scom:07} we have $a \in \xi'(m(p'))$. This means
that we have shown that $\xi(m(q')) \subset \xi(m(q'))$. As a consequence we have $m(p') < m(q')$. This proves one of the
implications of \ref{form:impl01}. Suppose now that $m(p') < m(q')$, which implies $\xi(m(q')) \subset \xi(m(p'))$. Consider $a \in
\xi'(q')$. Form \ref{scom:07} follows that $n(a) \in \xi(m(q'))$ and hence also $n(a) \in \xi(m(p'))$. Again from
\ref{scom:07} follows that $a \in \xi'(p')$. So we have shown that $\xi'(q') \subset \xi'(p')$, and as a consequence $p' < q'$.
Formula \ref{form:impl02} is proven in a completely analogous way. \qed

\section{Experiments And Preparations}
Some contexts are used to perform an experiment on the entity under consideration and other context are used to prepare a state of the
entity. We want to study these types of context more carefully, because they will play an important role in further operational foundations
of the formalism. Let us analyze what requirements are to be fulfilled for a context to be an experiment. 

For a context that is an experiment the context will change under influence of the state in such a
way that from the new context we can determine the outcome of the experiment. What are then outcomes of an experiment?

\subsection{Outcomes of Experiments}

For a context to play the role of an experiment it must be possible to identify outcomes of the experiment. We will denote outcomes
by
$x, y, z, \ldots$, and the set of possible outcomes corresponding to an experiment $e \in {\cal M}$, the entity being in
state
$p
\in \Sigma$, by $O(e, p)$. Obviously, the set of all possible outcomes of the experiment $e$ is then given by $\cup_{p \in
\Sigma}O(e, p) = O(e)$. 

If we consider the experimental practice in different scientific domains there does not seem to be a standard
way to identify outcomes. In general the description of an outcome of an experiment $e$ on an entity $S$
in state $p$ is linked to the state of the entity after the effect of the context, hence to the type of change that has been
provoked by the experiment, and also to the experiment itself, and to the new context that arises after the experiment has been performed.

This means that, if we consider a context
$e$ that we want to use as an experiment, and suppose that the entity is in state $p$, and that $q$ is a possible state that the entity can
change to under context
$e$, and $f$ is a possible context after $e$ has worked on $p$, then a possible outcome $x(f, q, e, p)$ for $e$ will occur.  But it might
well be that another possible state
$r$ that the entity might evolve to under context $e$, identifies the
same outcome
$x(f, r, e, p)= x(f, q, e, p)$ for
$e$. This is the case when state
$q$ only differs from state $r$ in aspects that are not relevant for the physical quantity measured by the experiment $e$. Let us give an
example to explain what we mean.

Suppose that we consider a classical physics entity $S$ that is a point particle located motionless
on a line that we have coordinated by the set of real numbers $\real$. The state of the particle in classical physics is described by its
position $u$ and its momentum $mv$, where $m$ is its mass and $v$ its velocity, hence by the vector $(u,mv) \in \real^2$. Our
experiment
$e$ consists of making a picture of the particle. On the picture we can read off the coordinate where the particle is, hence its
position. The set of possible outcomes
$O(e)$ for this experiment is a part of the set of real numbers
$\real$, namely the points described by the coordinate $u$, the position coordinate of the particle. The context $e$ is all what takes place
when we make the picture, but without the picture being taken, which means that for $e$ the film in the camera has not been exposed. This
context $e$ changes after the experiment into $f$ where the film has been exposed. The experiment $e$ is just an observation, not
provoking any change on the state. If for example outcome $x
\in
\real$ occurs, we know that the position $u$ of the particle equals $x$. The experiment not only
does not change the state of the particle, it is also a deterministic context. Indeed $O(e,(u,mv)) = \{u\}$ for all states $(u,mv)
\in \real^2$. In this example the experiment gathers knowledge about the state of the entity that we did not have before we
performed the experiment.

Let us consider the example of a one dimensional quantum particle, described by a state that is a wave function $\psi(x)$, element of
$L^2(\real)$ such that 
\[
\int\|\psi(x)\|^2dx = 1
\]
The context related to a position measurement is described by a self-adjoint operator with a set of spectral
projection operators that are the characteristic functions $\chi_\Omega$ of subsets $\Omega \subset \real$. The
probability for the quantum particle to be located in the subset
$\Omega \subset
\real$ by the effect of the context is given by 
\[
\int_{\Omega}\|\psi(x)\|^2dx
\]
and, if the particle is located in the subset $\Omega$, the state $\psi(x)$ is changed into
the state 
\be
{1 \over \sqrt{\int_\Omega\|\psi(x)\|^2dx}} \cdot \chi_\Omega \circ \psi (x) \label{eq:projstate}
\ee
In this new state, given by (\ref{eq:projstate}), the probability to be located in $\Omega$, if the position context is applied again to the
quantum particle, is given by:
\[
\int \|{1 \over \sqrt{\int_\Omega\|\psi(x)\|^2dx}} \cdot \chi_\Omega \circ \psi (x) \|^2 dx = 1
\]
That is the reason that we can consider the subset $\Omega$ as an outcome for the position experiment.
\bd [Experiment]
Consider a state context property system $(\Sigma, {\cal M}, {\cal L}, \mu, \xi)$ describing an entity $S$. We say that a context
$e \in {\cal M}$ is an experiment if for $p \in \Sigma$ there exists a set $O(e, p)$ and a map:
\bea
x: {\cal M} \times \Sigma \times {\cal M} \times \Sigma &\rightarrow& O(e, p) \\
(f, q, e, p) &\mapsto& x(f, q, e, p)
\eea
where $x(f, q, e, p)$ is the outcome of experiment $e$ for the entity in state $p$, and where $(e, p)$ has changed to $(f, q)$. We further
have that:
\be
\mu(f, q, e, p) \not= \{0\}
\ee
expressing that an outcome for $e$ is only possible if the transition probability from $(e, p)$ to $(f, q)$ is different from $\{0\}$. We
denote $q = P_x^e(p)$. The probability for the experiment $e$ to give outcome $x$ if the entity is in state $p$ equals 
\be
\mu(e, P_x^e(p), e, p)
\ee
For $x \in O(e, p) \cap O(e, r)$ we have:
\be
P_x^e(p) = P_x^e(r)
\ee
\ed
\noindent
The definition of an experiment that we have given here is still very general. As we saw already, in classical physics an experiment is
usually an observation, which is a much more specific type of experiment. An observation just `observes' the state of the entity that
is there without provoking any kind of change. Also in quantum mechanics an experiment is much less general
than the definition that we have given here. 

\subsection{Contexts and Experiments of the First Kind}

In quantum mechanics an experiment provokes a change of state such that the state after
the experiment is an eigenstate of this experiment, and the outcome is identified by means of this eigenstate. Experiments with this property
have been called experiments of the first kind in physics. 

\bd [Contexts of the First Kind]
Consider a state context property system $(\Sigma, {\cal M}, {\cal L}, \mu, \xi)$ describing an entity $S$. We say that a context $e \in {\cal
M}$ is a context of the first kind if for $p \in \Sigma$, we have that $(e, p)$ changes to $(f, q)$, with $f \in {\cal M}$ and $q \in \Sigma$,
where $q$ is an eigenstate of $f$.
\ed
\noindent
This means that for a context of the first kind, when $f$ is being applied again and again to the entity, its state will remain $q$. This
means that the entity has been changed into a stable state that no longer changes under influence of context. This is a perfect situation to
be able to identify an outcome of an experiment by means of this eigenstate.

\bd [Experiment of the First Kind]
Consider a state context property system $(\Sigma, {\cal M}, {\cal L}, \mu, \xi)$ describing an entity $S$. We say that $e \in {\cal M}$ is an
experiment of the first kind if $e$ is an experiment and a context of the first kind. This means that for $p \in \Sigma$ we have that
$P_x^e(p)$ is an eigenstate of $e$, in the sense that $\mu(e, P_x^e(p), e, P_x^e(p)) = \{1\}$, and the experiment $e$ makes occur the outcome
$x$ with probability equal to 1.
\ed
\noindent
An experiment of the first kind pushes each state of the entity into an eigenstate of this experiment. This is the way that experiments act in
quantum mechanics. Let us consider again the example of the experiment that measures the position of a quantum particle. The state $\psi(x)$
is changed to the state 
\[
{1 \over \sqrt{\int_\Omega\|\psi(x)\|^2dx}} \cdot \chi_\Omega \circ \psi(x)
\]
if we test whether the outcome is the interval $\Omega \subset \real$. If we test again whether the outcome is in the subset $\Omega$, the state
does not change any longer, because
\[
\chi_\Omega( {1 \over \sqrt{\int_\Omega\|\psi(x)\|^2dx}} \cdot \chi_\Omega \circ \psi(x)  ) = {1 \over \sqrt{\int_\Omega\|\psi(x)\|^2dx}}
\cdot \chi_\Omega \circ \psi(x)
\]
The set of outcomes for a quantum entity has the structure of the set of all subsets of another set, this other set being the spectrum of the
self-adjoint operator that represents the experiment in quantum mechanics. But there is more. Let us consider two consecutive
measurements of position, once in the subset $\Omega_1$ and second in the subset $\Omega_2$, such that
$\Omega_2 \subset \Omega_1 \subset \real$.
First, for $\Omega_1$ the state $\psi(x)$ changes to
\[
{1 \over \sqrt{\int_{\Omega_1}\|\psi(x)\|^2dx}} \cdot \chi_{\Omega_1} \circ \psi(x)
\]
with probability
\[
\int_{\Omega_1}\|\psi(x)\|^2dx
\]
Remark first that:
\[
\int_{\Omega_2} \|{1 \over \sqrt{\int_{\Omega_1}\|\psi(x)\|^2dx}} \cdot \chi_{\Omega_1} \circ \psi(x)\|^2 dx = {\int_{\Omega_2}\|\psi(x)\|^2dx
\over \int_{\Omega_1}\|\psi(x)\|^2dx}
\]
This means that for the second change of state the probability equals to:
\[
{\int_{\Omega_2}\|\psi(x)\|^2dx
\over \int_{\Omega_1}\|\psi(x)\|^2dx}
\]
and the state
\[
{1 \over \sqrt{\int_{\Omega_1}\|\psi(x)\|^2dx}} \cdot \chi_{\Omega_1} \circ \psi(x)
\]
is changed to
\[
\sqrt{{\int_{\Omega_1}\|\psi(x)\|^2dx 
\over \int_{\Omega_2}\|\psi(x)\|^2dx}} \cdot \chi_{\Omega_2} \circ {1 \over \sqrt{\int_{\Omega_1}\|\psi(x)\|^2dx}} \cdot \chi_{\Omega_1} \circ
\psi(x)
\]
Taking into account that $\chi_{\Omega_2} \circ \chi_{\Omega_1} \circ \psi(x) = \chi_{\Omega_2} \circ \psi(x)$ because $\Omega_2 \subset
\Omega_1$ we have that the final state is:
\[
{1 
\over \int_{\Omega_2}\|\psi(x)\|^2dx} \cdot \chi_{\Omega_2} \circ \psi(x)
\]
This is exactly the state that we would have found if we would immediately have measured the position with a test to see that the quantum
particle is localized in the subset $\Omega_2$. And also the probabilities multiply, namely the probability to find the position in
subset
$\Omega_2$ with a direct test on
$\Omega_2$ is the product of the probability to find it in $\Omega_1$ with a test on $\Omega_1$, with the probability to find it in
$\Omega_2$ after it had already been tested for $\Omega_1$. The process that we have identified here for the position measurement of a
quantum entity is generally true for all quantum mechanical measurements. Quantum mechanical measurement have a kind of cascade structure.
Let us introduce this structure for a general experiment and call it a cascade experiment. To be able to do so we need to define what we
mean by the product of two subsets of the interval $[0, 1]$ and also what we mean by 1 minus this subset for a subset of $[0, 1]$.
\bd
Suppose that $A, B, C \in {\cal P}([0, 1])$ of the interval $[0, 1]$. We define:
\bea
1 - A &=& \{1 - x\ \vert\ x \in A\} \\
B \cdot C &=& \{x \cdot y\ \vert\ x \in B, y \in C\}
\eea
\ed 
\noindent
Obviously $1 - A \in {\cal P}([0, 1])$ whenever $A \in {\cal P}([0, 1])$ and $A \cdot B \in {\cal P}([0, 1])$ whenever $A, B \in {\cal
P}([0, 1])$.
\bd [Cascade Experiment]
Consider a state context property system $(\Sigma, {\cal M}, {\cal L}, \mu, \xi)$ describing an entity $S$. An experiment $e \in {\cal M}$ is
a cascade experiment if there exists a set $E$ such that the set of outcomes $O(e)$ of $e$ is a subset ${\cal P}(E)$, hence $O(e) \subset
{\cal P}(E)$. For
$p
\in
\Sigma$, and
$x, y, z, t
\in O(e)$ such that
$x
\subset y$ and
$z
\cup t = E$ and $z \cap t = \emptyset$, we have:
\bea
P_y^e(P_x^e(p)) &=& P_x^e(p) \\
\mu(e, P_x^e(p), e, P_x^e(p)) &=& \{1\} \label{eq:idempotent} \\
\mu(e, P_x(p), e, p) &=& \mu(e, P_x(p), e, P_y(p)) \cdot \mu(e, P_y(p), e, p) \label{eq:cascadeexperiment} \\
\mu(e, P_z(p), e, p) &=& 1 - \mu(e, P_t(p), e, p)
\eea
We call $E$ the spectrum of the experiment $e$.
\ed
\noindent
Note that the elements of the spectrum $E$ for an experiment $e$ are not necessarily outcomes of the experiment $e$. Indeed, for example for
the position measurement of a free quantum particle, the spectrum of the position operator is a subset of the set of real numbers $\real$,
namely the spectrum of the self-adjoint operator corresponding to this position measurement, but none of the numbers of the spectrum is an
outcome. Only subsets of this spectrum with measure different from zero are outcomes, because the spectrum is continuous.

\subsection{An Extra Condition For the Morphisms}
When some of the contexts are experiments we can derive from the merological covariance situation an extra condition to be fulfilled for
the morphisms of ${\bf SCOP}$.

\bd
Consider two state context property systems $(\Sigma, {\cal M}, {\cal L}, \mu)$ and $(\Sigma', {\cal M}', {\cal L}', \mu')$. If $e \in
{\cal M}$ is an experiment, then also $l(e) \in {\cal M}'$ is an experiment, and for $p' \in \Sigma'$ there exists a bijection $k$
\be
k: O(e, m(p')) \rightarrow O(l(e), p')
\ee
which expresses that we use the same outcomes whether we experiment on the big entity $S'$ or on the subentity $S$.
\ed

\subsection{Preparations}
Context are also used to prepare the state of an entity. For a context to function as a preparation it is necessary that it brings
the entity in a specific state under its influence. 

\bd [Preparation]
Consider a state context property system $(\Sigma, {\cal M}, {\cal L}, \mu, \xi)$ describing an entity $S$. We say that a context
$e \in {\cal M}$ is a preparation if there exists a state $p \in \Sigma$ such that $R(e) = \{p\}$. We call $p$ the state prepared
by the context $e$.
\ed
\noindent
So a context is a preparation if it provokes a change such that each state of the entity is brought to one and the same state. This
is then the prepared state.

\section{Meet Properties and Join States} \label{sec:meetjoin}

Suppose we consider a set of
properties $(a_i)_i \in {\cal L}$. It is very well possible that there exist
states of the entity $S$ in which all the properties $a_i$ are
actual. This is in fact always the case if $\cap_i\kappa(a_i) \not=
\emptyset$. Indeed, if we consider $p \in \cap_i\kappa(a_i)$ and $S$
in state $p$, then all the properties $a_i$ are actual. If there corresponds a new property with the situation where all properties
$a_i$ of a set $(a_i)_i$ and no other are actual, we will denote such a new property by $\wedge_ia_i$,
and call it a `meet property' of all $a_i$. Clearly we have
$\wedge_ia_i$ is actual for $S$ in state $p \in \Sigma$ iff $a_i$ is
actual for all $i$ for $S$ in state $p$. This means that we have
$\wedge_ia_i \in \xi(p)$ iff $a_i \in \xi(p) \ \forall i$. 

Suppose now that we consider a set of states $(p_j)_j \in \Sigma$ of the entity
$S$. It is very well possible that there exist properties of the
entity such that these properties are actual if $S$ is in any one of
the states $p_j$. This is in fact always the case if $\cap_j\xi(p_j)
\not=
\emptyset$. Indeed suppose that $a \in \cap_j\xi(p_j)$. Then we have
that $a \in \xi(p_j)$ for each one of the states $p_j$, which means
that $a$ is actual if $S$ is in any one of the states $p_j$. If it
is such that there corresponds a new state to the situation where $S$ is in any one of the states
$p_j$, we will denote such new state by
$\vee_jp_j$ and call a `join state' of all $p_j$. We
can see that a property $a \in {\cal L}$ is actual for $S$ in
a state $\vee_jp_j$ iff this property $a$ is actual for $S$ in any of
the states $p_j$\footnote{A join state and meet property are not unique. But two join states and two meet properties corresponding
to the same sets are equivalent. We remark that we could also try to introduce join properties and meet states. It is however a
subtle but deep property of reality, that this cannot be done on the same level. We
will understand this better when we study in the next section more of the
operational aspects of the formalism. We will see there that
only meet properties and join states can be operationally defined in
the general situation.}. 
The existence of meet properties and join states gives
additional structure to $\Sigma$ and ${\cal L}$.

\bd [Property Completeness] \label{comstatprop}
Consider a state context property system $(\Sigma, {\cal M}, {\cal L}, \mu, \xi)$ describing an entity $S$. We
say that we have `property completeness' iff for an
arbitrary set $(a_i)_i, a_i \in {\cal L}$ of properties there exists
a property
$\wedge_ia_i \in {\cal L}$ such that for an
arbitrary state $p \in \Sigma$:
\begin{equation} \label{meetprop}
\wedge_ia_i \in \xi(p) \Leftrightarrow a_i \in \xi(p)\ \forall\ i
\end{equation}
Such a property $\wedge_ia_i$ is called a meet property of the set of properties $(a_i)_i$.
\ed
\bd [State Completeness]
Consider a state context property system $(\Sigma, {\cal M}, {\cal L}, \mu, \xi)$ describing an entity $S$. We say that we
have `state completeness' iff for an arbitrary set of states $(p_j)_j, p_j \in \Sigma$ there exists a state
$\vee_jp_j \in \Sigma$ such that for an
arbitrary property $a \in {\cal L}$:
\begin{equation} \label{joinstat}
\vee_jp_j \in \kappa(a) \Leftrightarrow p_j \in \kappa(a)\ \forall\ j
\end{equation}
Such a state $\vee_jp_j$ is called a join state of the set of states $(p_j)_j$.
\ed
\noindent
The following definition explains why we have introduced the concept completeness.
\bd [Complete Pre-ordered Set]
Suppose that $Z,<$ is a pre-ordered set. We say that
$Z$ is a complete pre-ordered set iff for each subset $(x_i)_i, x_i
\in Z$ of elements of $Z$ there exists an infimum and a
supremum in $Z$\footnote{An infimum of a subset
$(x_i)_i$ of a pre-ordered set $Z$ is an element of $Z$ that is smaller
than all the
$x_i$ and greater than any element that is smaller than all $x_i$. A
supremum of a subset $(x_i)_i$ of a pre-ordered set $Z$ is an element
of $Z$ that is greater than all the $x_i$ and smaller than any element
that is greater than all the $x_i$.}.
\ed
\bp
Consider a state context property system $(\Sigma, {\cal M}, {\cal L}, \mu, \xi)$ describing an entity $S$, and suppose that
we have property completeness and state completeness. Then
$\Sigma,
<$ and ${\cal L}, <$ are complete pre-ordered sets.
\ep
\noindent
Proof: Consider an arbitrary set $(a_i)_i, a_i \in {\cal L}$. We
will show that $\wedge_ia_i$ is an infimum. First we have to proof that
$\wedge_ia_i < a_k\ \forall\ k$. This follows immediately from
(\ref{meetprop}) and the definition of $<$ given in
(\ref{ordprop}). Indeed, from this definition follows that we have to
prove that $\kappa(\wedge_ia_i) \subset \kappa(a_k)\ \forall\ k$.
Consider
$p \in \kappa(\wedge_ia_i)$. From (\ref{statprop}) follows that this
implies that $\wedge_ia_i \in \xi(p)$. Through (\ref{meetprop}) this
implies that $a_k \in \xi(p)\ \forall\ k$. If we apply 
(\ref{statprop}) again this proves that $p \in \kappa(a_k)\ \forall\
k$. So we have shown that $\kappa(\wedge_ia_i) \subset \kappa(a_k)\
\forall\ k$. This shows already that $\wedge_ia_i$ is a lower bound
for the set $(a_i)_i$. Let us now show that it is a greatest lower
bound. So consider another lower bound, a property $b \in
{\cal L}$ such that $b < a_k\ \forall\ k$. Let us show that $b 
< \wedge_ia_i$. Consider $p \in \kappa(b)$, then we have $p \in
a_k\ \forall\ k$ since $b$ is a lower bound. This gives us that
$a_k \in \xi(p)\ \forall\ k$, and as a consequence $\wedge_ia_i \in
\xi(p)$. But this shows that $p \in \kappa(\wedge_ia_i)$. So we have
proven that $b < \wedge_ia_i$ and hence $\wedge_ia_i$ is an
infimum of the subset $(a_i)_i$. Let us now prove that $\vee_jp_j$ is
a supremum of the subset $(p_j)_j$. The proof is very similar, but we
use (\ref{joinstat}) in stead of (\ref{meetprop}). Let us again first
show that $\vee_jp_j$ is an upper bound of the subset $(p_j)_j$. We
have to show that $p_l < \vee_jp_j\ \forall\ l$. This
means that we have to prove that $\xi(\vee_jp_j) \subset \xi(p_l)\
\forall\ l$. Consider $a \in \xi(\vee_jp_j)$, then we have $\vee_jp_j
\in \kappa(a)$. From (\ref{joinstat}) it follows that $p_l \in
\kappa(a)\ \forall\ l$. As a consequence, and applying
(\ref{statprop}), we have that $a \in \xi(p_l)\ \forall\ l$. Let is now
prove that it is a least upper bound. Hence consider another upper
bound, meaning a state $q$, such that $p_l < q\ \forall\ l$. This
means that $\xi(q) \subset \xi(p_l)\ \forall\ l$. Consider now $a \in
\xi(q)$, then we have $a \in \xi(p_l)\ \forall\ l$. Using again
(\ref{statprop}), we have $p_l \in \kappa(a)\ \forall\ l$. From
(\ref{joinstat}) follows then that $\vee_jp_j \in \kappa(a)$ and
hence $a \in \xi(\vee_ja_j)$.

We have shown now that $\wedge_ia_i$ is
an infimum for the set $(a_i)_i, a_i \in {\cal L}$, and that
$\vee_jp_j$ is a supremum for the set $(p_j)_j, p_j \in \Sigma$.
It is a
mathematical consequence that for each subset $(a_i)_i, a_i \in {\cal
L}$, there exists also a supremum in ${\cal L}$, let is denote it by
$\vee_ia_i$, and that for each subset $(p_j)_j, p_j \in \Sigma$, there
exists also an infimum in
$\Sigma$, let us denote it by $\wedge_jp_j$. They are respectively
given by
$\vee_ia_i = \wedge_{x \in {\cal L}, a_i \prec x \forall i}\ x$ and
$\wedge_jp_j = \vee_{y \in \Sigma, y \prec p_j\forall j}\
y$\footnote{We remark that the supremum for elements of ${\cal L}$
and the infimum for elements of $\Sigma$, although they exists, as we
have proven here, have no simple operational meaning.}. \qed

\par
\medskip
\noindent
For
both ${\cal L}$ and $\Sigma$ it can be shown that this implies that
there is at least one minimal and one maximal element. Indeed, an
infimum of all elements of ${\cal L}$ is a minimal element of ${\cal
L}$ and an infimum of the empty set is
a maximal element of ${\cal L}$. In an
analogous way a supremum of all elements of $\Sigma$ is a maximal
element of
$\Sigma$ and a supremum of the empty set
is a minimal element of $\Sigma$. Of
course there can be more minimal and maximal elements. If a property
$a \in {\cal L}$ is minimal we will express this by $a \approx 0$ and
if a property $b \in {\cal L}$ is maximal we will express this by $b
\approx I$. An analogous notation will be used for the maximal and
minimal states.

When there is property completeness and state completeness we can specify the structure of the
maps $\xi$ and $\kappa$ somewhat more after having introduced the
concept of `property state' and `state property'.

\bp
Consider a state context property system $(\Sigma, {\cal M}, {\cal L}, \mu, \xi)$ describing an entity $S$, and suppose that
we have property completeness and state completeness. For $p \in \Sigma$ we define the `property state'
corresponding to $p$ as the property $s(p) = \wedge_{a \in
\xi(p)}a$. For $a \in {\cal L}$ we define the `state property'
corresponding to $a$ as the state $t(a) = \vee_{p \in \kappa(a)}p$.
We have two maps :
\begin{equation}
\begin{array}{ll}
t : {\cal L} \rightarrow \Sigma & a \mapsto t(a) \\
s : \Sigma \rightarrow {\cal L} & p \mapsto s(p)
\end{array}
\end{equation} 
and for $a, b \in {\cal L}$, and $(a_i)_i, a_i \in {\cal L}$ and
$p, q \in \Sigma$ and $(p_j)_j, p_j \in \Sigma$ we have :
\begin{equation}
\begin{array}{l}
a < b \Leftrightarrow t(a) < t(b) \\
p < q \Leftrightarrow s(p) < s(q) \\
t(\wedge_ia_i) \approx \wedge_it(a_i) \\
s(\vee_jp_j) \approx \vee_js(p_j)
\end{array}
\end{equation}
\ep
\noindent
Proof: Suppose that $p < q$. Then we have $\xi(q) \subset
\xi(p)$. From this follows that $s(p) = \wedge_{a \in \xi(p)}a <
\wedge_{a \in \xi(q)}a = s(q)$. Suppose now that $s(p) < s(q)$.
Take $a \in \xi(q)$, then we have $s(q) < a$. Hence also $s(p)
< a$. But this implies that $a \in \xi(p)$. Hence this shows that
$\xi(q) \subset \xi(p)$ and as a consequence we have $p < q$.
Because $\wedge_ia_i < a_k\ \forall\ k$ we have $t(\wedge_ia_i)
< t(a_k)\ \forall k$. This shows that $t(\wedge_ia_i)$ is a lower
bound for the set $(t(a_i))_i$. Let us show that it is a greatest
lower bound.
Suppose that $p < t(a_k)\ \forall\ k$. We remark that $t(a_k)
\in \kappa(a_k)$. Then it follows that $p \in \kappa(a_k)\ \forall\
k$. As a consequence we have $a_k \in \xi(p)\ \forall\ k$. But then
$\wedge_ia_i \in \xi(p)$ which shows that $p \in
\kappa(\wedge_ia_i)$. This proves that $p < t(\wedge_ia_i)$. So
we have shown that $t(\wedge_ia_i)$ is a greatest lower bound and
hence it is equivalent to $\wedge_it(a_i)$. \qed

\bp \label{interval}
Consider a state context property system $(\Sigma, {\cal M}, {\cal L}, \mu, \xi)$ describing an entity $S$, and suppose that
we have property completeness and state completeness. For $p \in \Sigma$ we have $\xi(p)
= [s(p), +\infty] = \{ a \in {\cal L} \
\vert \ s(p) < a\}$. For
$a
\in {\cal L}$ we have $\kappa(a) = [-\infty,
t(a)] = \{p \in \Sigma\ \vert\ p < t(a)\}$.
\ep
\noindent
Proof: Consider
$b \in [s(p), +\infty]$. This means that $s(p) < b$, and hence $b
\in
\xi(p)$. Consider now $b
\in \xi(p)$. Then $s(p) < b$ and hence $b \in [s(p), +\infty]$. \qed

\par
\medskip
\noindent
If $p$ is a state such that $\xi(p) = \emptyset$, this means that
there is no property actual for the entity being in state $p$. We
will call such states `improper' states. Hence a `proper' state is a
state that makes at least one property actual. In an analogous way,
if $\kappa(a) = \emptyset$, this means that there is no state that
makes the property $a$ actual. Such a property will be called an
`improper' property. A `proper' property is a property that is actual
for at least one state.

\bd [Proper States and Properties]
Consider a state context property system $(\Sigma, {\cal M}, {\cal L}, \mu, \xi)$ describing an entity $S$. We call $p \in
\Sigma$ a `proper' state iff $\xi(p) \not=
\emptyset$. We call $a \in {\cal L}$ a `proper' property iff
$\kappa(a) \not= \emptyset$. A state $p \in \Sigma$ such that $\xi(p)
= \emptyset$ is called an `improper' state, and a property $a \in {\cal
L}$ such that $\kappa(a) = \emptyset$ is called an `improper'
property. 
\ed
\noindent
It easily follows from proposition \ref{interval} that when there is property completeness and state completeness there are no
improper states ($I
\approx
\wedge\emptyset
\in
\xi(p)$) and no improper properties ($0 \approx \vee \emptyset \in
\kappa(a)$). Let us find out how the morphism behave in relation with `meet' and `join'.

\bp
Consider two state context property systems $(\Sigma, {\cal M}, {\cal L}, \mu, \xi)$ and $(\Sigma', {\cal M}', {\cal L}', \mu',
\xi')$ describing entities $S$ and $S'$ that are property and state complete, and a morphism $(m, l, n)$ between $(\Sigma, {\cal M},
{\cal L},
\mu,
\xi)$ and \\
$(\Sigma', {\cal M}', {\cal L}', \mu',
\xi')$. For $(a_i)_i \in {\cal L}$ and $(p'_j)_j \in \Sigma'$ we have:
\bea
n(\wedge_ia_i) &\approx& \wedge_in(a_i) \label{form:meet}\\
m(\vee_jp'_j) &\approx& \vee_jm(p'_j) \label{form:join}
\eea
\ep
\noindent
Proof: Because $\wedge_ia_i < a_j\ \forall j$ we have $n(\wedge_ia_i) < n(a_j)\ \forall j$. From this follows that $n(\wedge_ia_i)
< \wedge_in(a_i)$. There remains to prove that $\wedge_in(a_i) < n(\wedge_ia_i)$. Suppose that $\wedge_in(a_i) \in \xi'(p')$. Then
$n(a_j) \in \xi'(p')\ \forall j$, which implies that $a_j \in \xi(m(p'))\ \forall j$. As a consequence we have $\wedge_ia_i \in
\xi(m(p'))$. From this follows that $n(\wedge_ia_i) \in \xi'(p')$. So we have proven that $\wedge_in(a_i) < n(\wedge_ia_i)$.
Formula \ref{form:join} is proven in an analogous way. \qed

\section{Operationality}

We have introduced states, contexts and properties for a physical entity. In this section we analyze in which way
operationality introduces connections between these concepts.

\subsection{Testing Properties and Operationality}

Experiments can be used to measure many things, and in this sense they can also be used to test properties. Let us explain how this
works. Consider an experiment $e \in {\cal M}$ and a property $a \in {\cal L}$. If there exists a subset $A \subset
O(e)$ of the outcome set of $e$, such that the property $a$ is actual iff the outcome of $e$ is contained in $A$, we say that $e$
tests the property $a$.

\bd [Test of a Property]
Consider a state context property system $(\Sigma, {\cal M}, {\cal L}, \mu, \xi)$ describing an entity $S$. If for a property $a \in
{\cal L}$ there is an experiment $e \in {\cal M}$, and a subset $A \subset O(e)$ of the outcome set of $e$, such that:
\be
a \in \xi(p) \Leftrightarrow O(e, p) \subset A
\ee
We say that $e$ is a `test' for the property $a$.
\ed
\noindent
If all the properties of the entity that we consider can be tested by an experiment, we say that we have operationality, or that
our entity is an operational entity.
\bd [Operational Entity]
Consider a state context property system $(\Sigma, {\cal M}, {\cal L}, \mu, \xi)$ describing an entity $S$. If for each property $a
\in {\cal L}$ there is an experiment $e \in {\cal M}$ that tests this property, and if moreover the experiments to test the
properties are such that for two properties $a, b \in {\cal L}$ we have at least experiments $e, f \in {\cal M}$ that test
respectively $a$ and $b$ such that $O(e) \cap O(f) = \emptyset$, we say that the entity $S$ is an operational entity.
\ed
\noindent
Of course, for an operational entity, it is not necessary to give the set of properties apart, they can be derived from the rest of the
mathematical structure. We have done this explicitly in \cite{Aerts2002} for the case where all experiment contexts are yes/no-experiments.
It is possible to generalize the construction of \cite{Aerts2002}. For this reason we need to introduce what we will call a state context
system.

\bd [The Category SCO]
A state context system $(\Sigma, {\cal M}, \mu)$ consists of two sets $\Sigma$ and ${\cal M}$, and a function
\be
\mu: {\cal M} \times \Sigma \times {\cal M} \times \Sigma \rightarrow {\cal P}([0, 1])
\ee
The sets $\Sigma$ and ${\cal M}$ play the role of the set of states and the set of contexts of an entity $S$, and the function $\mu$
describes the transition probability. Consider two state context systems $(\Sigma, {\cal M}, \mu)$ and $(\Sigma', {\cal M}', \mu')$. A
morphism is a couple of functions $(m, l)$ such that:
\bea
m: \Sigma' &\rightarrow& \Sigma \\
l: {\cal M} &\rightarrow& {\cal M}'
\eea
and the following formula is satisfied:
\be
\mu(f, m(q'), e, m(p')) = \mu(l(f), q', l(e), p')
\ee
Further we have that if $e \in {\cal M}$ is an experiment, then also $l(e) \in {\cal M}'$ is an experiment and for $p' \in \Sigma'$ we have
a bijection $k$:
\be
k: O(e, m(p')) \rightarrow O(l(e), p') \label{eq:bijectionoutcomes}
\ee
We denote the category of state context systems and their morphisms by ${\bf SCO}$.
\ed
\noindent
For such a state context system we can construct the set of properties that are testable by experiments of ${\cal M}$, and this will
deliver us a state context property system for an operational entity. Let us see how this works.

\bd 
Suppose that we have a state context system $(\Sigma, {\cal M}, \mu)$. We define:
\bea
{\cal L} &=& \{A\ \vert\ \exists\ e \in {\cal M},\ e\ {\rm experiment\ and}\ A \subset O(e)\} \\
\xi&:& \Sigma \rightarrow {\cal P}({\cal L}) \\
A &\in& \xi(p) \Leftrightarrow O(e, p) \subset A
\eea
and call $(\Sigma, {\cal M}, {\cal L}, \mu, \xi)$ the state context property system related to $(\Sigma, {\cal M}, \mu)$.
\ed

\bp
Suppose that $(\Sigma, {\cal M}, {\cal L}, \mu, \xi)$ and $(\Sigma', {\cal M}', {\cal L}', \mu', \xi')$ are the state context property
systems related to the state context systems $(\Sigma, {\cal M}, \mu)$ and $(\Sigma', {\cal M}', \mu)'$. A morphism $(m, l)$ between
$(\Sigma, {\cal M}, \mu)$ and $(\Sigma', {\cal M}', \mu)'$  determines a morphism $(m, l, n)$ between
$(\Sigma, {\cal M}, {\cal L}, \mu, \xi)$ and $(\Sigma', {\cal M}', {\cal L}', \mu', \xi')$.
\ep
\noindent
Proof: Consider $A \in {\cal L}$. This means that $\exists\ e \in {\cal M}$ where $e$ is an experiment, and $A \subset O(e)$. We know that
$l(e) \in {\cal M}'$ is also an experiment, and if we consider $k(A)$, where $k$ is the bijection of (\ref{eq:bijectionoutcomes}) we have
$k(A)
\subset O(l(e)) = k(O(e))$. This means that
$k(A)
\in {\cal L}'$. Let us define:
\bea
n: {\cal L} &\rightarrow& {\cal L}' \\
A &\mapsto& k(A)
\eea
Take $p' \in \Sigma'$ and $A \in {\cal L}$. We have $A \in \xi(m(p')) \Leftrightarrow O(e, m(p')) \subset A \Leftrightarrow k(O(e, m(p')))
\subset k(A) \Leftrightarrow O(l(e), p') \subset n(A) \Leftrightarrow n(A) \in \xi'(p')$. This means that $(m, l, n)$ is a morphism between 
$(\Sigma, {\cal M}, {\cal L}, \mu, \xi)$ and $(\Sigma', {\cal M}', {\cal L}', \mu', \xi')$.

\medskip
\noindent
We introduce in the next section specific types of contexts and states that make it possible to test the meet property and
deliver a join state for our entity.

\subsection{Product Contexts and Product States}
Suppose that we consider a set of contexts $(e_i)_i$. In general it will be possible for only one of these contexts to be
realized together with an entity $S$. We can however consider the following operation: we choose one of the contexts of the
set $(e_i)_i$ and realize this context together with the entity $S$. We can consider this operation together with the set of
contexts $(e_i)_i$ as a new context. Let us denote it as $\Pi_ie_i$ and call it the product context of the set $(e_i)_i$. It
is interesting to note that the product of different contexts gives rise to indeterminism. In earlier work we have been
able to prove that the quantum type of indeterminism is exactly due to the fact that each experiment is the product of some hidden
experiments. We have called this explanation of the quantum probability structure the `hidden measurement'
approach\cite{Aerts1986,Aerts1987,Aerts1994}. In \cite{Aerts2002} we analyze in detail how we have to introduce the product
experiment in a mathematical way, and it is shown how a subset probability is necessary for this purpose.

\bd [Product Context]
Consider a state context property system $(\Sigma, {\cal M}, {\cal L}, \mu, \xi)$ describing an entity $S$. Suppose we have a set
of contexts $(e_i)_i \in {\cal M}$. The product context $\Pi_ie_i$ is defined in the following way. For $p, q \in \Sigma$ and $f
\in {\cal M}$ we have:
\be
\mu(f, q, \Pi_ie_i, p) = \cup_i\mu(f, q, e_i, p) \label{eq:productcontext} \\
\ee
\ed 
\bp
Consider a state context property system $(\Sigma, {\cal M}, {\cal L}, \mu, \xi)$ describing an entity $S$. For the product context
$\Pi_ie_i$ of a set of contexts $(e_i)_i \in {\cal M}$ we have for $p \in \Sigma$:
\be
R(\Pi_ie_i, p) = \cup_iR(e_i, p) \label{eq:productcontext01}
\ee
\ep
\noindent
Proof: Suppose now that $q \in R(\Pi_ie_i, p)$. This means that there exist $f \in {\cal M}$ such that $\mu(f, q, \Pi_ie_i, p) \not=
\{0\}$. Hence $\cup_i\mu(f, q, e_i, p) \not= \{0\}$, which means that there is at least $j$ such that $\mu(f, q, e_j, p) \not=
\{0\}$. This shows that $q \in R(e_j, p)$, and hence $q \in \cup_iR(e_i, p)$. On the contrary, suppose that $q \in \cup_iR(e_i,
p)$. This means that there is at least one $j$ such that $q \in R(e_j, p)$. Hence there exist $f \in {\cal M}$ such that $\mu(f, q,
e_j, p) \not= \{0\}$. As a consequence we have $\cup_i\mu(f, q, e_i, p) \not= \{0\}$, and hence $\mu(f, q, \Pi_ie_i, p) \not=
\{0\}$, which shows that $q \in R(\Pi_ie_i, p)$. This proves (\ref{eq:productcontext01}).\qed

\bp
Consider a state context property system $(\Sigma, {\cal M}, {\cal L}, \mu, \xi)$ describing an entity $S$. Suppose that $(e_i)_i$
is a set of experiments. We have:
\bea
O(\Pi_ie_i, p) = \cup_iO(e_i, p) \label{eq:productcontext02}
\eea
\ep
\noindent
Proof: Suppose that $x \in O(\Pi_ie_i, p)$. This means that there exist $q \in \Sigma$ and $f \in {\cal M}$ such that $\mu(f, q,
\Pi_ie_i, p) \not= \{0\}$, and $x$ occurs whenever $p$ changes into $q$. Because of (\ref{eq:productcontext}) we have $\cup_i\mu(f,
q, e_i, p) \not= \{0\}$. This means that there is at least one $j$ such that $\mu(f, q, e_j, p) \not= \{0\}$. Hence $x$ is a
possible outcome of $e_j$ that occurs when $p$ is changed into $q$ by $e_j$. As a consequence we have $x \in O(e_j, p)$, and hence
$x \in \cup_iO(e_i, p)$. On the contrary, suppose now that $x \in \cup_iO(e_i, p)$. This means that there is at least one $j$ such
that $x \in O(e_j, p)$. This means that there exist $q \in \Sigma$ and $f \in {\cal M}$ such that $\mu(f, q, e_j, p) \not= \{0\}$,
and $x$ is the outcome that occurs when $e_j$ changes the state $p$ to the state $q$. As a consequence we have $\cup_i\mu(f, q, e_i,
p) \not= \{0\}$, which means that $\mu(f, q,
\Pi_ie_i, p) \not= \{0\}$, which means that $x$ also occurs when $\Pi_ie_i$ changes the state $p$ to $q$. Hence $x \in O(\Pi_ie_i,
p)$. This proves (\ref{eq:productcontext02}). \qed

\medskip
\noindent
Suppose that we
consider a set of states $(p_i)_i \in \Sigma$. Then it is possible to consider a situation where the entity is in one of these
states, but we do not know which one, as a new state, that we will call the product state $\Pi_ip_i$ of the set of states
$(p_i)_i$. More specifically we define the product state as the state that is prepared by a product of contexts that are
preparations.
\bd [Product State]
Consider a state context property system $(\Sigma, {\cal M}, {\cal L}, \mu, \xi)$ describing an entity $S$. Suppose that
$(p_i)_i
\in \Sigma$ is a set of states. The product state $\Pi_ip_i$ is defined in the following way, for $e, f \in {\cal M}$ and $q \in
\Sigma$ we have:
\be
\mu(f, q, e, \Pi_ip_i) = \cup_i\mu(f, q, e, p_i)
\ee
\ed
\bp
Consider a state context property system $(\Sigma, {\cal M}, {\cal L}, \mu, \xi)$ describing an entity $S$. For the product state
$\Pi_ip_i$ of a set of states $(p_i)_i \in \Sigma$ we have for $e \in {\cal M}$:
\be
R(e, \Pi_ip_i) = \cup_iR(e, p_i)
\ee
\ep
\noindent
Proof: Suppose that $q \in R(e, \Pi_ip_i)$. This means that there exists $f \in {\cal M}$ such that $\mu(f, q, e, \Pi_ip_i) \not=
\{0\}$. Hence $\cup_i\mu(f, q, e, p_i) \not= \{0\}$. This means that there is at least one $j$ such that $\mu(f, q, e, p_j) \not=
\{0\}$. Hence $q \in R(e, p_j)$ which shows that $q \in \cup_iR(e, p_i)$. On the contrary, suppose now that $q \in \cup_iR(e,
p_i)$. This means that there is at least one $j$ such that $q \in R(e, p_j)$. Hence there exists $f \in {\cal M}$ such that $\mu(f,
q, e, p_j) \not= \{0\}$. As a consequence we have $\cup_i\mu(f, q, e, p_j) \not= \{0\}$, and hence $\mu(f, q, e, \Pi_ip_i) \not=
\{0\}$. This shows that $q \in R(e, \Pi_ip_i)$. \qed

\subsection{Meet Properties and Join States}
We can show that product contexts test meet properties while product states are join states.

\bp
Consider a state context property system $(\Sigma, {\cal M}, {\cal L}, \mu, \xi)$ describing an entity $S$. Consider a set of
properties
$(a_i)_i \in {\cal L}$. Suppose that we have experiments $(e_i)_i$ available, such that experiment $e_j$ tests property $a_j$, and
such that $O(e_j) \cap O(e_k) = \emptyset$ for $j \not= k$, then the product experiment $\Pi_ie_i$ tests the meet property
$\wedge_ia_i$.
\ep
\noindent
Proof: Experiment $e_i$ tests property $a_i$. This means there exists $A_i \subset O(e_i)$ such that $a_i \in \xi(p) \Leftrightarrow O(e, p) \subset
A_i$. Consider now the product experiment
$\Pi_ie_i$.  We will prove that $\Pi_ie_i$ tests the property
$\wedge_ia_i$. Consider $A = \cup_iA_i$. Then we have $\cup_iA_i \subset \cup_iO(e_i) = O(\Pi_ie_i)$. We have $O(\Pi_ie_i, p) =
\cup_iO(e_i, p)$ and, since $O(e_i) \cap O(e_j) = \emptyset$ for $i \not= j$ we also have $O(e_i, p) \cap O(e_j, p) = \emptyset$
for $i \not= j$. This means that $\cup_iO(e_i, p) \subset \cup_iA_i \Leftrightarrow O(e_j, p) \subset A_j\ \forall j$. Consider the
property $a$ tested as follows by $\Pi_ie_i$: $a \in \xi(p) \Leftrightarrow O(\Pi_ie_i, p) \subset A$. Then $a \in \xi(p)
\Leftrightarrow a_j \in \xi(p)\ \forall j$, which proves that $a = \wedge_ia_i$. \qed

\bp
Consider a state context property system $(\Sigma, {\cal M}, {\cal L}, \mu, \xi)$ describing an operational entity $S$. Suppose
that we have a set of states $(p_i)_i \in \Sigma$, then the product state $\Pi_ip_i$ is a join state of the set of states $(p_i)_i$.
\ep
\noindent
Proof: Suppose that we have $\Pi_ip_i \in \kappa(a)$ for $a \in {\cal L}$. Since the entity is operational we have an experiment
$e \in {\cal M}$ that tests the property
$a$. This means that there exists $A \subset O(e)$ such that $a \in \xi(p) \Leftrightarrow O(e, p) \subset A$. From $\Pi_ip_i \in
\kappa(a)$ follows that $a \in \xi(\Pi_ip_i)$. Hence $O(e, \Pi_ip_i) \subset A$. As a consequence we have $\cup_iO(e, p_i) \subset
A$, which shows that $O(e, p_j) \subset A\ \forall j$, and hence $a \in \xi(p_j)\ \forall j$. From this follows that $p_j \in
\kappa(a)\ \forall j$. On the contrary, suppose now that $p_j \in \kappa(a)\ \forall j$, where $e$ is again an experiment that
tests the property $a \in {\cal L}$. Then there exists $A \subset O(e)$ such that $O(e, p_j) \subset A\ \forall j$. As a
consequence we have that $\cup_iO(e, p_i) \subset A$. Hence $O(e, \Pi_ip_i) \subset A$. From this follows that $\Pi_ip_i \in
\kappa(a)$. This means that we have proven that $\Pi_ip_i$ is a join state of the set of states $(p_i)_i$. \qed

\bd [Operational Completeness]
Consider a state context property system $(\Sigma, {\cal M}, {\cal L}, \mu, \xi)$ describing an operational entity $S$. We will say
that the state context property system is operationally complete if for any set $(e_i)_i \in {\cal M}$ of contexts the product
context $\Pi_ie_i \in {\cal M}$ and for any set of states $(p_i)_i$ the product state $\Pi_ip_i \in \Sigma$.
\ed

\begin{theorem}
Any operationally complete state context property system \\ $(\Sigma, {\cal M}, {\cal L}, \mu, \xi)$ satisfies property completeness
and state completeness. 
\end{theorem}
Let us introduce the category with elements the operationally complete state context property systems.

\section{Conclusion}
There remains a lot of work to make the formalism that we put forward into a full grown theory. Potentially however such a theory
will be able to describe dynamical change as well as change by a measurement in a unified way. Both are considered to be contextual change.
Certainly for application in other fields of reality this generality will be of value. As for applications to physics it will be
interesting to reconsider the quantum axiomatics and reformulate the essential axioms within the formalism that is proposed here. This
project has been elaborated already within the earlier axiomatic approaches, which means that part of the work is translation into the more
general scheme that we present here. However, because the basic concepts are different it will not just be a translation of earlier
results. In the years to come we plan to engage in this enterprise.

\end{document}